\documentclass{biophys-new}
\usepackage[utf8]{inputenc}
\usepackage{graphicx}
\usepackage{float}
\usepackage[colorlinks,allcolors=cyan!70!black]{hyperref}
\usepackage{xcolor}

\usepackage{lipsum}
\usepackage{ulem}

\title{Viscoelastic phenotyping of red blood cells}
\runningtitle{Biophysical Journal Template} 

\author[1,2,*]{M. Gironella-Torrent}
\author[3]{G. Bergamaschi} 
\author[4]{R. Sorkin}
\author[3]{G. Wuite}
\author[1,5]{F. Ritort}

\runningauthor{Author1 and Author2} 

\affil[1]{Small Biosystems Lab, Condensed Matter Physics Department, University of Barcelona, 08028 Barcelona, Spain}
\affil[2]{Department of Medical Biochemistry and Cell Biology, Institute of Biomedicine, The Sahlgrenska Academy, University of Gothenburg, 40530 Gothenburg, Sweden}
\affil[3]{Department of Physics and Astronomy and LaserLab, Vrije Universiteit Amsterdam, 1081 HV Amsterdam, The Netherlands}
\affil[4]{Raymond and Beverly Sackler Faculty of Exact Sciences, School of Chemistry, Tel Aviv University, Tel Aviv, Israel}
\affil[5]{Institut de Nanoci\`encia i Nanotecnologia (IN2UB), Universitat de Barcelona, 08028 Barcelona, Spain}

\corrauthor[*]{ritort@ub.edu}

\papertype{Article}

\begin{document}

\begin{frontmatter}

\begin{abstract}
Red Blood Cells (RBCs) are the simplest cell types with complex dynamical and viscoelastic phenomenology. While the mechanical rigidity and the flickering noise of RBCs have been extensively investigated, an accurate determination of the constitutive equations of the relaxational kinetics is lacking. Here we measure the force relaxation of RBCs under different types of tensional and compressive extension-jump protocols by attaching an optically trapped bead to the RBC membrane. Relaxational kinetics follows linear response from 60pN (tensional) to -20pN (compressive) applied forces, exhibiting a triple-exponential function with three well-separated timescales over four decades (0.01-100s). While the fast timescale ($\tau_F\sim 0.02(1)s$) corresponds to the relaxation of the membrane, the intermediate and slow timescales ($\tau_I=4(1)s$; $\tau_S=70(8)s$) likely arise from the cortex dynamics and the cytosol viscosity. Relaxation is highly heterogeneous across the RBC population, yet the three relaxation times are correlated, showing dynamical scaling. Finally, we find that glucose depletion and laser illumination of RBCs lead to faster triple-exponential kinetics and RBC rigidification. Viscoelastic phenotyping is a promising dynamical biomarker applicable to other cell types and active systems. 
\end{abstract}

\begin{sigstatement}
This research shows the structured viscoelastic dynamics of red blood cells (RBCs) and highlights the significance of considering multiple timescales for understanding their mechanical behavior. The observed triple-exponential relaxation behavior, coupled with the proposed viscoelastic model, provides valuable insights into the underlying processes governing RBC mechanics. Furthermore, our findings regarding the impact of glucose depletion and light illumination on RBC rigidity show how environmental factors affect RBC properties. Our results expand the current knowledge of RBC mechanics and pave the way for future investigations of relaxational phenomena in other cell types.
\end{sigstatement}
\end{frontmatter}

\section*{Introduction}

Red blood cells (RBCs) are the most abundant and simplest cells in the human body. They are transported through the bloodstream to deliver oxygen to the body tissues. While they circulate through capillaries, RBCs are subject to mechanical deformation and stress. Capillaries can be thin as half the RBC disk diameter (6-8$\mu$m), causing RBCs to squeeze as they pass through. Mechanical properties of RBCs are tightly related to shape and composition, which are crucial for oxygen transport and delivery. They are also highly dependent on endogenous (e.g., genetic) and exogenous (e.g., physicochemical stresses, aging) factors and are determinants for homeostasis. If altered, they lead to diseases and disorders, such as hemolytic anemias and thrombosis. Therefore, RBC mechanical properties are key biomarkers for human health.

 In humans, mature RBCs lack a nucleus to maximize the storage of hemoglobin and oxygen transport capacity. Moreover, the typical biconcave shape of RBCs increases their surface area facilitating oxygen diffusion \cite{dean2005abo}. The lifespan of human RBCs is about 115 days \cite{franco2012measurement}. During this period, they cyclically circulate in the bloodstream lasting for about a minute per cycle \cite{blom2003monitoring} and are subject to continuous deformation. 

 Mechanical deformability dysfunction is directly related to diseases such as sickle anemia \cite{nash1984mechanical}, malaria \cite{mauritz2010biophotonic}, and thalassemia \cite{vaya2015association}. Moreover, during their lifespan, RBCs suffer from numerous age-dependent alterations that form the RBC aging phenotype, such as a decline of metabolic activity, cell shape modification \cite{dmitrieff2017balance}, oxidative injury \cite{inanc2021quantifying}, mechanical fatigue \cite{qiang2019mechanical}, among others \cite{antonelou2010aging}. These alterations trigger erythrophagocytosis or the ingestion of RBCs by macrophages \cite{arias2017red}.
It has been reported that most RBC dysfunctions lead to a rigidification of the cell \cite{nash1984mechanical,mauritz2010biophotonic,vaya2015association,embury1984concurrent,liu2019mechanical}. 

RBCs are viscoelastic, showing complex time-dependent responses to external perturbations. Dynamical biomarkers such as frequency-dependent elastic moduli yield valuable information about their physiological state. In the past, the viscoelastic response of living cells and RBCs has been studied with techniques as diverse as micropipette aspiration \cite{artmann1997micropipette,wang2022fluorescence,chien1978theoretical}, deformation in a flow \cite{fedosov2014deformation,cranston1984plasmodium,artmann1995microscopic}, AFM \cite{ciasca2015mapping}, acoustic force spectroscopy (AFS) \cite{sorkin2018probing}, optical magnetic twisting cytometry (MTC) \cite{puig2007viscoelasticity} and laser optical tweezers (LOT) \cite{zhu2020optical,yoon2008nonlinear,mills2004nonlinear,henon1999new}. These techniques have also been employed to study active polymer networks \cite{head2003deformation,mizuno2007nonequilibrium}. The dynamical response of RBCs depends on the type of perturbation applied, the geometry of the experimental configuration, and the measured physicochemical property. Mechanical stiffness measurements have reported values that change by two orders of magnitude depending on the pulling orientation, the bond attachment type, and the probe's contact area. Noise correlation spectroscopy has emerged as an excellent technique to investigate rheological phenomena and activity of RBCs, e.g., using LOT \cite{brochard1975frequency,turlier2019unveiling,yoon2011red,betz2009atp,evans2017geometric}. It is useful for measuring membrane fluctuations (flickering) in the high-frequency domain (sub-second timescale). In the low-frequency domain ($\ge $1s), active fluctuations and nonequilibrium phenomena are observed \cite{betz2009atp,mizuno2007nonequilibrium,turlier2016equilibrium}. RBCs flickering has been shown to depend on the viscosity of the surrounding medium, demonstrating that flickering is an active process \cite{tuvia1997cell}. The power spectrum of the flickering signal at low frequencies increases with ATP concentration contributing to an average entropy production of thousands of $k_B$T/s per $\mu$m$^2$ of surface area \cite{turlier2016equilibrium,di2023variance}. Therefore, a quantitative characterization of RBC relaxation in the low-frequency domain is key to a better understanding of RBC hemostasis \cite{cluitmans2016red}. 

Here we carry out a new class of tensional and compressive extension-jump experiments in RBC with optical tweezers and measure force relaxation over four decades (0.01-100s). We discover a previously unknown triple-exponential force relaxation with three well-separated timescales (fast, intermediate, and slow) that permits us to extract the constitutive parameters (stiffness and friction coefficient) of the dissipative processes in the RBC. The three timescales emerge from the underlying RBC architecture, providing a dynamical biomarker for RBC phenotyping based on viscoelastic relaxation experiments.  

\section*{Materials and Methods}

{\bf Sample preparation.} 
Human RBCs were freshly prepared before each experiment by finger-pricking of a healthy donor. Four microliters of blood were diluted in 1 mL of a PBS solution containing 130 mM NaCl, 20 mM K/Na phosphate buffer, 10 mM glucose, and 1 mg/mL BSA \cite{betz2009atp}. Two types of beads were used for the experiments: polystyrene beads for non-specific attachment (diameter 3$\mu$m) and silica beads coated with concanavalin-A (diameter 3$\mu$m) for the specific attachments to the RBC. 

{\bf Experimental setup.} 
Experiments were done in a miniaturized version of an optical tweezers instrument described in \cite{dieterich2015single}. Measurements were performed with highly stable miniaturized laser tweezers in the dual-trap mode that consists of two counter-propagating laser beams focused on the same spot creating an optical trap. The instrument directly measures forces by linear momentum conservation.  Piezo actuators bend the optical fibers and allow us to move the trap while measuring its position using a light lever that deflects 5$\%$ of the light to a position-sensitive detector (PSD). Force and trap position measurements are acquired at 1 kHz bandwidth. The experiments are performed in a microfluidics chamber made of two glass coverslips and two sheets of parafilm with three channels: the central one, where the optical trap and the micropipette are located; and two supplier channels, each one with a dispenser tube, to supply beads and RBCs. Details of the pipette and microfluidics chamber can be found in \cite{gieseler2021optical}.

{\bf Experimental configuration.} 
The RBC is vertically held between two micron-sized beads in the experimental configuration. One bead is immobilized on the tip of a micropipette by air suction, while the other is optically trapped. The pulling configuration requires the following steps. First, a bead from the first dispenser tube is captured in the central channel by the optical trap and moved close to the exit of the second dispenser tube where RBCs flow to the central channel. Second, a floating RBC in the central channel is brought in contact and attached to the optically trapped bead. A uniform laminar buffer flow is applied to the central channel to prevent RBCs from being optically trapped and damaged by the laser. Moreover, the flowed buffer cleans the working area from other RBCs that can interfere with the experiment. While applying the flow, the bead attached to the RBC is fixed to the tip of the micropipette. A second bead is captured in the trap and attached to the opposite extreme of the RBC. This second bead is used as the force probe for the experiments, Figure \ref{Fig:1}a. Further information can be found in \cite{gieseler2021optical}.

{\bf Pulling experiments.} 
For each RBC, a single force-distance curve is obtained. To compute the RBC stiffness, the relative trap position is converted into cell extension by subtracting the bead-trap displacement ($F/k_b$) from the measured trap-pipette distance with $k_b$ the trap stiffness \cite{dieterich2015single,gieseler2021optical}. Once the force-extension curve is measured, a linear fit between 7.5pN and 12.5pN is performed by using \textit{fit} function of MATLAB. For forces below 5 pN, we observe a certain variability in the force-extension curves (FEC) due to the positioning of the contact point between the RBC and the lower bead. Significantly, this variation does not arise from the RBC itself but is a consequence of the distinct geometry in each experimental configuration. On the contrary, for higher forces (approximately 20 pN), tether formation becomes evident, which in turn has an impact on the FEC. This complexity makes it difficult to directly compare the results across different cases, as illustrated in Figure \ref{Fig:1}e. Therefore, the force range between 7.5pN and 12.5pN provides suitable conditions where variability is minimized.

{\bf Force-relaxation fits.} Force-time relaxation curves were obtained by averaging force and time values in a logarithmic scale.  The logarithmic averaged data is fitted to Eq. \ref{eq:3exp} with MATLAB using the \textit{fit} function by imposing boundaries to the fitting parameters to minimize the computational time. The lower boundary for relaxation times is 0s, and the upper boundary is 500s. Regarding amplitudes, the lower boundary is -$\Delta F$ and the upper boundary is $\Delta F$, with $\Delta F$ the magnitude of the force jump. The stationary force fitting parameter has no boundaries. Fitting parameters with their uncertainty range are obtained using the \textit{confint} function in MATLAB with a $68 \%$ of confidence.

\section*{Results}
In the experiments, a single RBC is attached to two polystyrene beads ($3\mu m$ diameter), each one at opposite ends of the cell (Figure \ref{Fig:1}a). The lower bead (LB) is immobilized by air suction on the tip of a micropipette (MP), whereas the upper bead (UB) is captured in the optical trap (OT). The experimental configuration requires a series of manual operations where beads are captured with the optical trap and attached to the RBC (see Materials and Methods). A flow is applied to the RBC to keep it far from the laser focus to avoid optical damage of the RBC by direct illumination \cite{zhu2020optical,inanc2021quantifying}. Two kinds of experiments have been performed: pulling and extension-jump experiments. In pulling experiments, the RBC is repeatedly stretched back and forth between a minimum and a maximum force, while the force-extension curve (FEC) is measured. In extension-jump experiments, one or more step-jumps are applied to the trap position (section \textit{Force-relaxation experiments}), and the force-relaxation curve (FRC) is measured.   

\subsection*{Pulling experiments}
\begin{figure*}[h]
\centering
\includegraphics[width=1\linewidth]{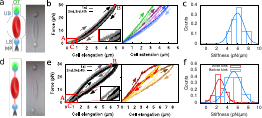}
\caption[Pulling experiments on untethered (U) and tethered (T) RBCs]{\textbf{Pulling experiments on untethered (U) and tethered (T) RBCs.} (\textbf{a}) Schematics and video image of our experimental configuration of the untethered population where the lower micro-sized bead (LB) is fixed on the tip of a micropipette (MP) and the upper bead (UB) is captured in the optical trap (OT) shown in green. (\textbf{b}) (Left panel) First four force-extension curves of an RBC without the tether. In grey, the first FEC shows a remanent elongation, and, in black, the three consecutive FECs overlap with each other. (Right panel) FECs for 3 different RBCs, each one represented with a different color, the pulling curve (from 0pN to 30 pN) is represented in dark color while the pushing curve (from 30 pN to 0pN) is represented in a light color. (\textbf{c}) Histogram of untethered RBC stiffness obtained from the FECs of 8 different RBCs. The stiffness is computed by performing a linear fit of the FEC between 7.5 pN and 12.5 pN. (\textbf{d}) Schematics and video images of the experimental configuration of the tethered population. In the image, the RBC presents two tethers, one attached to each bead, but it is also possible to have RBCs with only one tether in the lower or upper bead. (\textbf{e}) (Left panel) First four FECs of an RBC with tether. In grey, the first FEC presents a remanent elongation and, in black, the consecutive three FECs, which overlap with each other. (Right panel) Force-extension curves for 3 different RBCs (different colors) in the presence of tether. The dark color represents the pulling curve from 0pN to 30pN, and the light color represents the pushing curve from 30pN to 0pN. (\textbf{f}) Stiffness histograms for the pulling FECs in the presence of tether. In blue, the stiffness values before the kink and, in red, the values after the kink (indicated as a dot in panel e, right).}
\label{Fig:1}
\end{figure*}

Figure \ref{Fig:1}b (left) shows FECs of a single RBC repeatedly pulled between 0 and 30pN. The first stretch-release pulling cycle (left panel, grey) systematically deviates from the following ones (left panel, black) in agreement with previous results \cite{yoon2008nonlinear}. A fully relaxed RBC (point A) is mechanically stretched at $\sim$1pN/s up to 30pN (point B). Then the trap is moved backward at the same speed until 0pN, at which point a remanent elongation of $\sim 0.5 \mu$ m is observed (point C and zoom). We interpret the remanent elongation as the irreversible orientation of the RBC that occurs in the first pulling cycle \footnote{The same phenomenon is observed in ferromagnetic materials where the initial magnetization curve differs from the rest due to the irreversible motion of the domain walls \cite{chikazumi2009physics}}. Figure \ref{Fig:1}b (right panel) shows FECs for three different RBCs, which present mechanical hysteresis between the stretching (dark color) and releasing (light color) parts of the cycle (upward vs downward arrows). FECs show cell-to-cell variability indicating the heterogeneous mechanical response of the RBC population. From the FECs, we also extracted the (elongational) stiffness of the RBCs defined as the slope of the FEC between 7.5 and 12.5 pN (Fig \ref{Fig:1}c). The mean value $(6.1\pm1.3)$pN/$\mu$m is compatible with previous experimental and simulation results with LOT \cite{mills2004nonlinear}.

Sometimes, a tether forms at sufficiently high forces (around 25-30 pN). This observation is in accordance with the results obtained through micropipette aspiration \cite{waugh2001membrane}. This is shown in Figure \ref{Fig:1}d where two tethers, each at one side of the RBC, are extruded above $\simeq 20$pN \footnote{We have also observed cases where a single tether is extruded in one of the two sides only (indistinctly)}. Figure \ref{Fig:1}e (left) shows the first five pulling cycles of a single RBC in the presence of a tether. Like in the untethered case, a similar remanent elongation ($\sim 0.5 \mu$m) is observed after the first pulling cycle (Figure \ref{Fig:1}e left, zoom), while Figure \ref{Fig:1}e (right) shows FECs for three different RBCs with tether formation.

The FECs with tether show higher hysteresis than those without tether (Figure \ref{Fig:1}b and e, right). Tether formation is observed as the detachment of one of the two beads from the cell body at a given force.  Concomitantly, a kink appears in the FEC at that force where the stiffness drops. 
The same phenomenon is observed in the releasing process:  a kink appears in the FEC upon tether absorption by the cell membrane. RBCs with a tether also show cell-to-cell variability and heterogeneous mechanical response across the RBC population (Figure \ref{Fig:1}e, right). Tether extrusion and absorption are cooperative processes at characteristic forces (maroon and red dots respectively in Figure \ref{Fig:1}e, right). Tether formation is an irreversible process with hysteresis: tether extrusion occurs at forces higher than tether absorption (Figure \ref{Fig:S1}). 

We also carried out experiments using beads functionalized with concanavalin A, which specifically attaches to the RBC spectrin network, finding similar results. In particular, we also found the same spread in cell-to-cell variability as for the case of non-specific attachments, showing that heterogeneous mechanical response is an intrinsic feature of RBCs (Figure \ref{Fig:S2}).

\subsection*{Force-relaxation experiments}
\label{sec:FRC}
RBCs have a complex architecture consisting of two major structures: the membrane lipid bilayer and the 2D spectrin-actin network. One might identify structural contributions to the overall relaxation from mechanical stretching experiments. Here we combine different kinds of extension-jump protocols and identify three well-separated and reproducible timescales in RBCs over four decades (0.01-300 seconds). 

\begin{figure*}[h]
\centering
\includegraphics[width=1\linewidth]{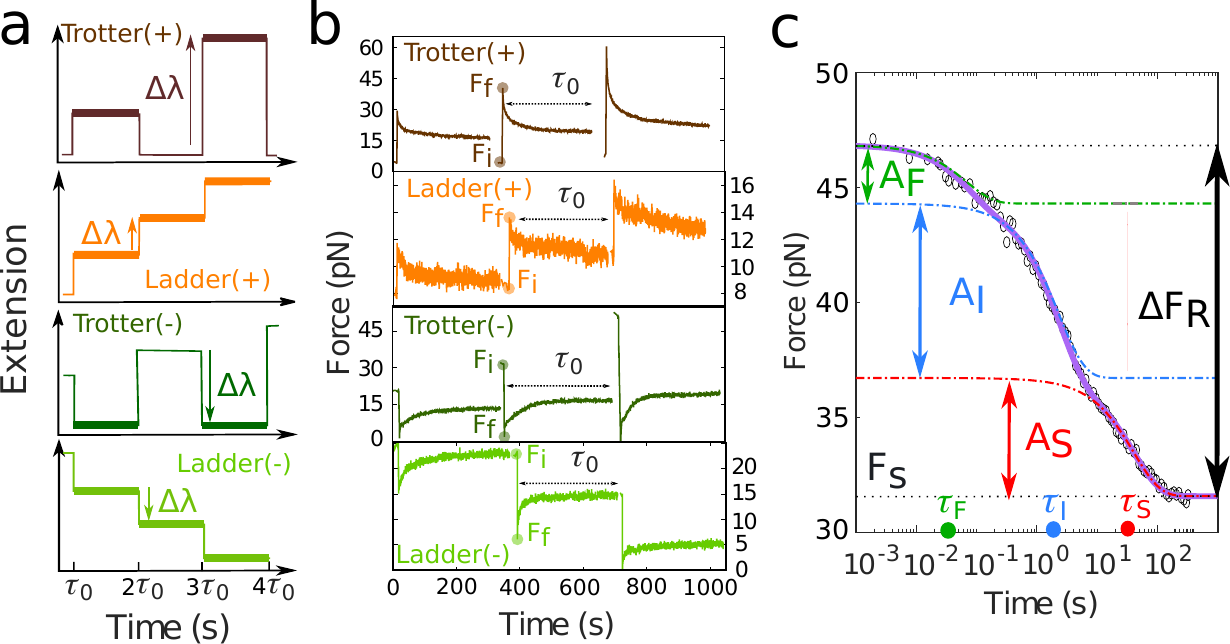}
\caption[Protocols, force-relaxation curves (FRCs) and triple-exponential fit.] { \textbf{Protocols, force-relaxation curves (FRCs), and triple-exponential fit.} (\textbf{a}) Schematic representation of trap displacement versus time for the different protocols (trotter(+), ladder(+), trotter(-), and ladder(-)). In thick lines, the time windows during which measurements are performed. The arrow next to $\Delta \lambda$ indicates the direction of the jump. (\textbf{b}) Three FRC for each of the four different protocols in (a). Initial and final forces are indicated with big dots in the second FRC of each panel, defining the force jump $\Delta F=F_f-F_i$. (\textbf{c}) FRC in normal-log scale. Data points were obtained by average filtering. The fitting curve Eq. \ref{eq:3exp} is shown in purple. Parameters are shown for the fast ($A_F$ and $\tau_F$, green); intermediate ($A_I$ and $\tau_I$, blue); and slow ($A_S$ and $\tau_S$, red) processes. The stationary force, $F_s$, is shown as a black dotted line at the bottom.}
\label{Fig:2}
\end{figure*}

Two kinds of step-wise extension-jump protocols have been implemented, trotter and ladder (Fig.\ref{Fig:2}a). Extension jumps are applied by instantaneously moving the trap position by $\Delta\lambda$ ($\sim 1-3\mu$m).  $\Delta\lambda>0$ for tensional jumps (Fig.\ref{Fig:2}a, upper panels) while $\Delta\lambda<0$ for compressive jumps (Fig.\ref{Fig:2}a, lower panels). The time between consecutive jumps is fixed to $\tau_0=5$ min (Fig.\ref{Fig:2}a, bottom axis). The trap displacement $\Delta\lambda$ causes a force jump, $\Delta F$ = $F_f - F_i$, where $F_i$ is the initial force before the jump, and $F_f$ is the force immediately after the jump (Fig.\ref{Fig:2}b). Force-relaxation curves (FRCs) are subsequently measured during $\tau_0=5$min (Fig.\ref{Fig:2}b). We label relaxation curves by $\Delta F$, being positive (negative) for tensional (compressive) jumps. In trotter-type protocols, a single extension jump is applied (Fig.\ref{Fig:2}a, brown and dark green), and FRCs measured (Fig.\ref{Fig:2}b, brown and dark green). The initial (final) force stays close to zero ($<5$pN) for tensional (compressive) jumps. In ladder-type protocols, extension jumps are consecutively applied (Fig.\ref{Fig:2} in orange and light green), and FRCs measured at every ladder step  (Fig.\ref{Fig:2}b, orange and light green). In all protocols, several jumps (between two and five) of varying amplitude ($\Delta F$) were applied, the total experimental time per RBC being roughly half an hour. FRCs in Fig.\ref{Fig:2}b show a viscoelastic response over minutes upon RBC deformation. For tensional jumps ($\Delta\lambda,\Delta F>0$, upper panels), force monotonically decreases with time, indicating RBC expansion. For compressive jumps ($\Delta\lambda,\Delta F<0$, lower panels), force monotonically increases with time, indicating RBC contraction. In all protocols, FRCs relax towards a stationary force $F_S$ that is higher (lower) than the initial force $F_i$ for tensional (compressive) protocols.

Since the pioneering studies by Evans et al. in 1976 \cite{evans1976membrane} on RBC membrane viscoelasticity, a significant volume of research has been dedicated to exploring the shear modulus, viscoelastic properties, and relaxation dynamics. Most of the literature on RBC relaxation \cite{cranston1984plasmodium,hochmuth1979red,henon1999new,artmann1995microscopic,qiang2019mechanical,bronkhorst1995new,dao2003mechanics,chien1978theoretical} reports that RBCs regain their shape in a time-dependent manner, exhibiting a characteristic single exponential behavior consistent with the viscoelastic solid model. It is worth noting that the tensile/compressive deformations can also be referred to as loading/relaxation phases. Remarkably, and in contrast with these studies, we observe that FRCs show a triple exponential behavior (Figure \ref{Fig:2}c),
\begin{equation}
    F(t) = A_F e^{-t/\tau_F} + A_I e^{-t/\tau_I} + A_S e^{-t/\tau_S} + F_s
\label{eq:3exp}
\end{equation}
where $F_s=F(t\to\infty)$ and the $A's$ and $\tau's$ define amplitudes and relaxation times of the fast $(A_F,\tau_F)$, intermediate $(A_I,\tau_I)$, and slow $(A_S,\tau_S)$ processes.

\begin{figure}[h]
\centering
\includegraphics[width=1\linewidth]{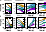}
\caption[Experimental force-relaxation curves (FRCs) for the four different protocols]{ \textbf{Experimental force-relaxation curves (FRCs) for the four different protocols.} Force versus time in normal-log scale. Each panel shows the raw data (different colors) and their corresponding triple-exponential fits Eq. \ref{eq:3exp} (black) for the four different protocols of Figure \ref{Fig:2}. Each color represents a relaxation curve with a different $\Delta F$. (\textbf{a}) For the untethered population and (\textbf{b}) for the tethered population.}
\label{Fig:3}
\end{figure}

In Figure \ref{Fig:2}c we illustrate the parameters of the triple exponential for one relaxation curve of the Trotter(+) type. 
A crucial feature of the FRC is the recovery force $\Delta F_R$, defined as the total force change during the relaxation. From Eq. \ref{eq:3exp}, $\Delta F_R=F(t=0)-F_s=A_F+A_I+A_S$ (Fig. \ref{Fig:2}c).  $\Delta F_R$ measures the viscoelastic response of the RBC after it has been deformed. Notice that $\Delta F_R = 0$ would indicate a purely elastic response. We find that $\Delta F_R$ is roughly $\Delta F/2$, being positive (negative) for tensional (compressive) deformations exhibiting a viscoelastic response. Viscoelastic behavior has been previously reported in RBCs \cite{mills2004nonlinear}, epithelial cells \cite{wei2008comparative}, and shape memory polymers \cite{lendlein2002shape}. Figure \ref{Fig:3} shows four selected FRCs with different protocols without a tether (panels a) and with tether extrusion (panels b), together with their fits to Eq. \ref{eq:3exp} (black curves). FRCs are qualitatively the same and a triple exponential is needed to fit the data in all cases. While other functions such as a double exponential and a stretched exponential fit the relaxational curves in some conditions, they do not consistently fit all the data (Figs. \ref{Fig:S3} and \ref{Fig:S7}).

\begin{figure*}[h]
\centering
\includegraphics[width=1.0\linewidth]{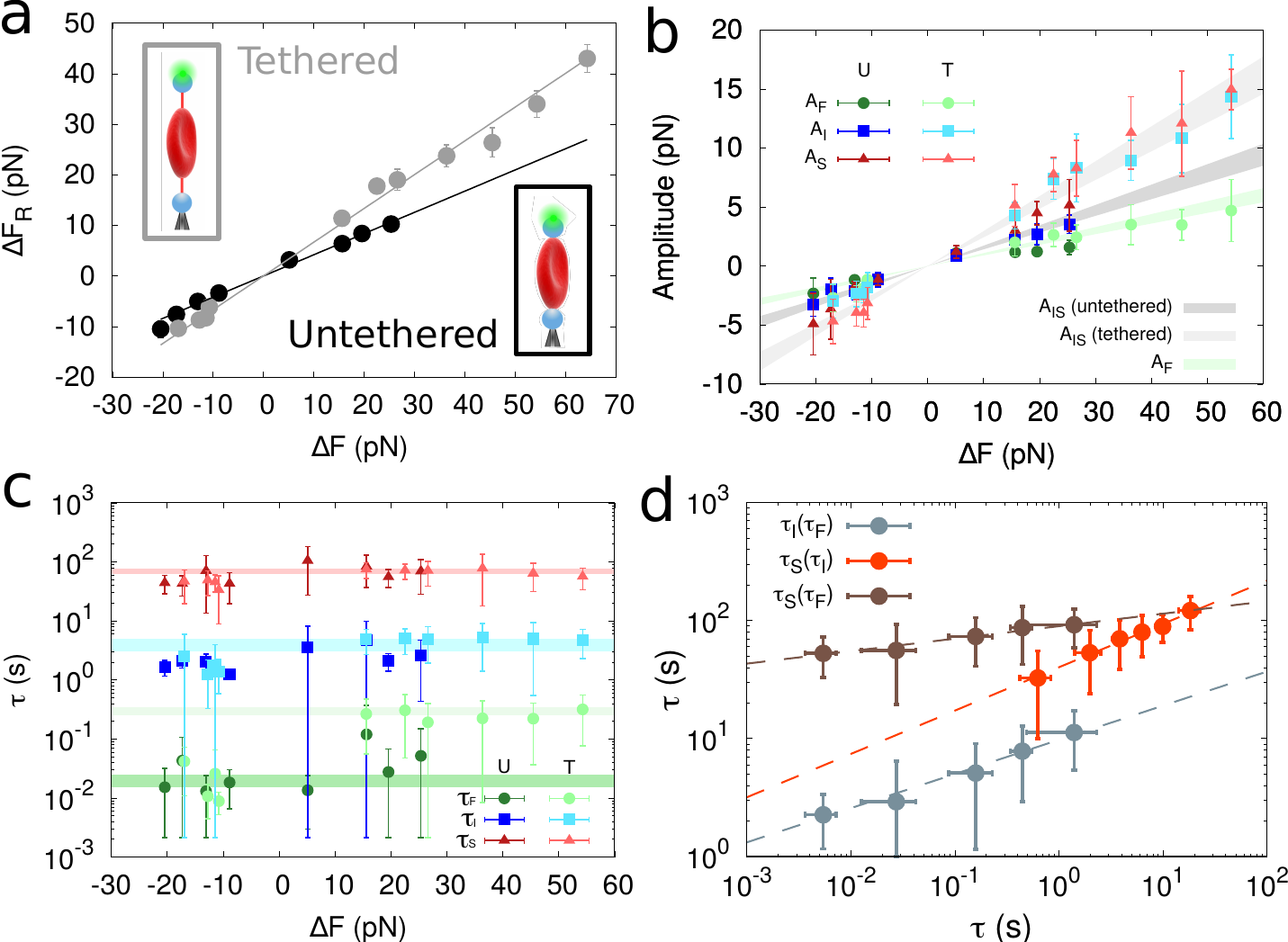}
\caption[Recovery forces, amplitudes and relaxation times ]{\textbf{Recovery forces, amplitudes and relaxation times .} (\textbf{a}) Recovery force $\Delta F_R$ versus $\Delta F$. In grey, the tethered case, and in black, the untethered case. (\textbf{b}) Amplitudes versus $\Delta F$ for both types of RBC populations. In dark color, the untethered case (U) and, in light color, the tethered case (T). Green represents the fast amplitude, blue represents the intermediate amplitude, and red represents the slow amplitude. (\textbf{c}) Characteristic times versus $\Delta F$ for both tethered and untethered cases. In dark color, the untethered (U) case and, in light color, the tethered (T) case. Green represents the fast time $\tau_F$, blue represents the intermediate time $\tau_I$, and red represents the slow time $\tau_S$. (\textbf{d}) Dynamical scaling for the three characteristic times taken over all experimental data (U and T). }
\label{Fig:4}
\end{figure*}

Figure \ref{Fig:4} shows all parameters of the fits (amplitudes and relaxation times) plotted versus the force jump $\Delta F$. Results have been averaged over the different protocols (Trotter(+) with Ladder (+) for $\Delta F>0$ and Trotter(-) with Ladder (-) for $\Delta F<0$). $\Delta F_R$ and the amplitudes $A_F,A_I,A_s$ show a linear dependence with $\Delta F$ (Figure \ref{Fig:4}a,b). The same result is obtained for each protocol independently (Figure \ref{Fig:S4}). At a closer inspection, fitting parameters show a systematic difference between untethered and tethered RBCs. Therefore, these are shown separately for each case, every point being the average over 3-6 RBCs. In Figure \ref{Fig:4}a, we plot $\Delta F_R$ versus $\Delta F$, untethered (U, black circles) and tethered (T, grey circles). A linear relation is found in both cases, $\Delta F_R=0.42(2)\Delta F$ (untethered) and $\Delta F_R=0.67(2)\Delta F$ (tethered) showing higher viscous response (i.e. larger $\Delta F_R$) for tethered RBCs. The higher viscosity of the extruded tether leads to the higher hysteresis observed in the FECs (Fig. \ref{Fig:1}e). Tether extrusion occurs around 25pN, so measurements for the untethered case were limited to force jumps $\Delta F<$ 25pN. Figure \ref{Fig:4}b shows the three amplitudes $A_F$ (green circles), $A_I$ (blue squares), $A_S$ (red triangles) plotted versus $\Delta F$ averaged over the different protocols. Again, a linear relation is found for the three amplitudes in the tethered (T, light color) and untethered (U, dark color) cases. Moreover, $A_I$ and $A_S$ (blue and red symbols) show similar linear slopes but differ from those for $A_F$ (green symbols). Indeed, amplitudes $A_I,A_S$ are compatible for the tethered and untethered cases. A single linear fit to $A_I$ and $A_S$ for all data gives $A_{IS}=0.27(3)\Delta F$ (tether, light grey area) and $A_{IS}=0.15(2)\Delta F$ (untethered, dark grey area). In contrast, for the fast amplitude $A_F$, similar values are obtained for the untethered (dark green) and tethered (light green) cases, $A_F=0.11(1)\Delta F$ (light green area). We hypothesize that the fast process arises from the viscoelastic response between the bead and the RBC at the local contact area, for both tethered and untethered RBCs. Instead, the difference in $A_{IS}$ between untethered and tethered suggests that the intermediate and slow processes depend on the different morphology of the RBC in the two cases (video images in Figure \ref{Fig:1}a,d). The fact $A_{\rm I}\simeq A_{\rm S}\equiv A_{\rm IS}$ is indicative of two equally stiff responsive viscoelastic elements connected in parallel (see next section).

Further information can be obtained from the relaxation times $\tau_F$, $\tau_I$, $\tau_S$. The results are shown in Fig. \ref{Fig:4}c. Remarkably, we find three distinct well-separated timescales independent of $\Delta F$ and the protocol. Moreover, these timescales are the same for tethered and untethered RBCs, except for $\tau_F$ for $\Delta F>0$, which is larger for the tethered case (light green versus dark green circles for $\Delta F>0$ in Fig. \ref{Fig:4}c). Tether extrusion ($\Delta F>0$) causes the RBC's membrane to flow from the RBC body toward the growing tether. The mass exchange process during the flow slows down the fast process by increasing $\tau_F$. In fact, upon retraction of the RBC ($\Delta F<0$), the tether is not absorbed, so $\tau_F$ remains unchanged for the untethered and tethered RBC (left most green points in Fig. \ref{Fig:4}c). Video images of the RBC's retraction show that the RBC's body does not fully absorb the tether, reducing the mass flow from the tether to the RBC. The  irreversible flow of the membrane upon tether extrusion and retraction leads to the asymmetry of $\tau_F$ between $\Delta F >0$ (movie S1, Sec.\ref{Movie}) and $\Delta F <0$ (movie S2, Sec.\ref{Movie}).  Averaging the values of $\tau_F$ over $\Delta F>0$, we get $\tau_F(\Delta F>0)=0.30(5)$s for the T case (light green band), whereas for the U case (all $\Delta F$) and T case ($\Delta F>0$) we get $\tau_F=0.020(5)$s (dark green band). By averaging $\tau_I$, $\tau_S$ over T, U and $\Delta F$, we get $\tau_I = \rm 4(1)$s (cyan band), $\tau_S = \rm 70(8)$s (light red band). While $\tau_F$ and $\tau_I$ can be linked to the relaxation of the cell membrane, $\tau_S$ corresponds to the relaxation of the RBC's body, which takes significantly more time.

Remarkably, timescales $\tau_F$, $\tau_I$, and $\tau_S$ exhibit a strong heterogeneity across the RBC population throughout the experimental data (U and T), varying over several orders of magnitude. Despite this variability, timescales are strongly correlated, showing dynamical scaling. Figure \ref{Fig:4}d shows the three timescales plotted relative to each other. Timescales are power-law correlated, a signature of dynamic scaling, $\tau_i=\alpha \tau_j^\beta$ with $i,j=F,I,S$ and $\alpha,\beta>0$. 
We obtain $\tau_I=10(1)\tau_F^{0.29(2)}$ (grey points), $\tau_S=40(2)\tau_I^{0.37(2)}$ (orange points), $\tau_S=89(3)\tau_F^{0.11(1)}$ (brown points).

\subsection*{Viscoleastic model}
\label{sec:model}
\begin{figure}[h]
\centering
\includegraphics[width=0.4\linewidth]{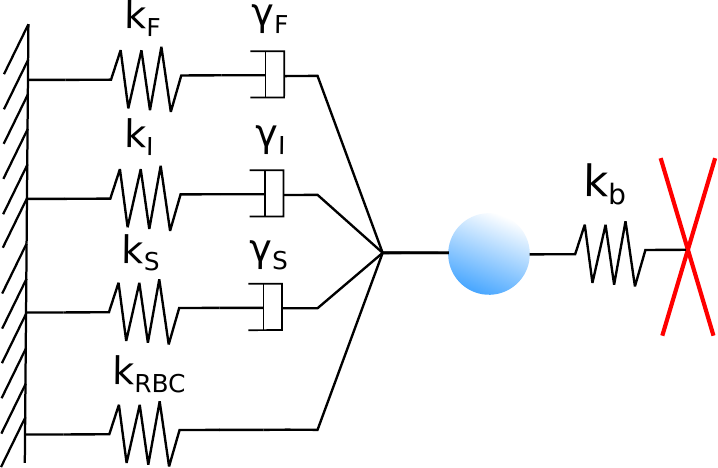}
\caption[]{\textbf{Schematics of the viscoelastic model.} The RBC contribution is represented in the left side of the bead with three spring-dashpot elements and a single spring element connected in parallel. The optical trap is modeled as a spring on the right side of the bead. For more details, see Sec.\ref{sec:S21Model} and Figure \ref{Fig:S5}.}
\label{Fig:5}
\end{figure}

The triple exponential relaxation Eq. \ref{eq:3exp} can be reproduced by a viscoelastic model that combines the bead in the optical trap of stiffness $k_b$ with a Maxwell-Wiechert configuration for the RBC. The latter consists of four elements: one spring of stiffness $k_{RBC}$ and three spring-dashpots of stiffness $k_F$,$k_I$,$k_S$ and friction coefficients $\gamma_F$, $\gamma_I$, $\gamma_S$ that are connected in parallel (Fig. \ref{Fig:5}). The model can be analytically solved (Supplemental Sec.\ref{sec:S21Model}) under the approximation of timescale separation $\tau_F\ll \tau_I\ll \tau_S$ as observed in the experiments. We get,

\begin{equation}
    \Delta F_R = \frac{\Delta F}{(1+\frac{k_{RBC}}{k_b})(1+\frac{k_{RBC}}{k_{||}})}
\label{eq:DFr}
\end{equation}

\begin{equation}
    A_i = \frac{k_i}{k_{RBC}+k_{||}} \Delta F
\label{eq:Ai}
\end{equation}

\begin{equation}
    \tau_i= \frac{k_{RBC}+k_b+k_{||}}{k_i(k_{RBC}+k_b+k_{||}-k_i)}\gamma_i
\label{eq:Taui}
\end{equation}
with $i\equiv F,I,S$ and $k_{||}=k_F+k_I+k_S$. Using the values of $k_{RBC}$ obtained in the pulling experiments (Fig.\ref{Fig:1}c,f), we can solve the equations and determine the six parameters $k_F,k_I,k_S,\gamma_F,\gamma_I,\gamma_S$ for the U and T cases. Results are summarized in Table \ref{table:tablevisco}. We remark on three results. First, $\gamma_F$ for the T case is ten times larger than for the U case, which we interpret as the additional friction caused by the tether on the RBC relaxation. Second, parameters for the I, and S  spring-dashpot elements are equal for the U and T cases, indicating that the I and S processes correspond to the cell body relaxation, which is unaffected by tether formation. While the stiffnesses $k_I,k_S$ are similar, friction coefficients $\gamma_I,\gamma_S$ differ by a factor of 20, being also 100-1000 times larger than $\gamma_F$. The high $\gamma_I,\gamma_S$ values explain the slow relaxation kinetics (up to tens of minutes) observed in the FRCs. This phenomenology conforms with the strong violations of the fluctuation-dissipation theorem observed at low frequencies in flickering noise experiments \cite{turlier2016equilibrium}.
\begin{table}[htp]
\centering
\begin{tabular}{c|ccccc}

${\rm pN}/\mu{\rm m}$   & $k_{RBC}$  &  $k_{||}$ & $k_F$ & $k_I$ & $k_S$ \\ 
\hline
$\textbf{U}$ & 6.1(13)  & 5.3(1)  &  1.3(1) & 2.0(1) & 2.0(1) \\
$\textbf{T}$ & 2.7(6)  & 6.4(4)  &  1.0(1) & 2.7(5) & 2.7(5)  \\
\hline\hline
${\rm pN}\cdot {\rm s/\mu m}$ &  $\gamma_F$ & $\gamma_I$   & $\gamma_S$   \\ 
\hline
$\textbf{U}$ & 0.025(3)  & 7.8(6) & 136(12) & \\
$\textbf{T}$ & 0.29(2)  & 10.2(7) & 179(13) & \\
 \hline
\end{tabular}
\bigskip
  \caption{Fitting parameters of the viscoelastic model. Parameters are obtained by fitting Eqs. \ref{eq:DFr},\ref{eq:Ai},\ref{eq:Taui} to the measured recovery force, amplitudes, and relaxation times.}
     \label{table:tablevisco}
\end{table}

\subsection*{Rigidification of treated and irradiated RBCs}
\begin{figure}[h]
\centering
\includegraphics[width=1.0\linewidth]{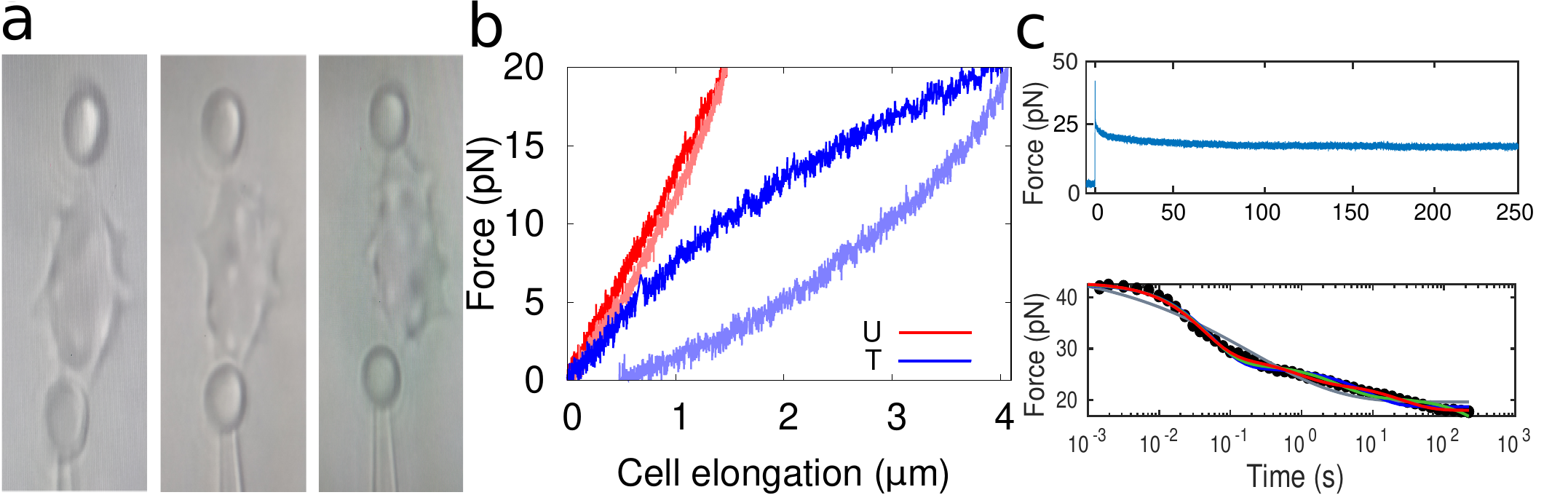}
\caption[ATP-depleted RBC.]{\textbf{ATP-depleted RBC.} (\textbf{a}) Three images of ATP-depleted RBC in echinocyte form. (\textbf{b}) Pulling cycles for a representative untethered (U, red) and tethered (T, blue) RBC. (\textbf{c},top panel) Typical force-relaxation curve. (\textbf{c},bottom panel) Relaxation curve together with its fits to a triple exponential (red), triple exponential with $A_I=A_S$ (green), double exponential (blue), and stretched exponential (grey). Results are shown for a representative RBC and are reproducible over different RBCs.}
\label{Fig:6}
\end{figure}

Under environmental damage, RBCs suffer structural modifications changing shape, and developing membrane instabilities. Figure \ref{Fig:6}a shows images of an RBC after partially depleting glucose and consequently ATP. RBCs were treated by incubating in a glucose-free buffer at 37 $^\circ$C for 20h (48h are needed for full depletion \cite{turlier2016equilibrium}). ATP is essential for RBC activity and membrane stability. ATP depletion prevents the dissociation/reassociation activity in the membrane-cortex, and the spectrin network is rigidified \cite{gov2005red,gov2007active,ben2011effective}. As shown in Figure \ref{Fig:6}a, RBCs develop membrane spikes with a star-like geometry (echinocyte). We have repeated the pulling experiments in this condition, finding  much less stable RBC-bead attachments and RBC rigidification in the U ($k_{RBC}=13(3)$pN/$\mu$m, Fig. \ref{Fig:S6}) and T ($k_{RBC}=6(1)$pN/$\mu$m) cases (Fig.\ref{Fig:6}b).  The values of $k_{RBC}$ in treated and untreated for the untethered case are compatible with previous results \cite{betz2009atp,rodriguez2015direct}. Interestingly, treated RBCs mostly exhibit tether formation in relaxation experiments, indicating membrane instability caused by changes in the relative area difference between the two leaflets of the lipid bilayer, which trigger the echinocyte transformation at low levels of ATP \cite{gov2005red}. Additionally, it has been reported \cite{borghi2007tether} that the ATP is crucial for the recovery of mechanical properties in RBCs after tether extrusion. This is because ATP depletion impedes complete retraction and promotes subsequent extrusion.  Again, all FRCs show a viscoelastic linear response with a triple exponential Eq. \ref{eq:3exp} (Figure \ref{Fig:6}c and \ref{Fig:S7}). We find a slightly higher viscoelastic response $\Delta F_R=0.71(2)\Delta F$ (Fig. \ref{Fig:S8}a) but lower intermediate and slow amplitudes ($A_{IS}=0.20(3)\Delta F$, Fig. \ref{Fig:S8}b) compared to the previously shown untreated T case ($A_{IS}=0.27(3)\Delta F$). In contrast, the fast amplitude triplicates its value ($A_{F}=0.31(3)\Delta F$) compared to the untreated T case ($A_{F}=0.11(1)\Delta F$). The relaxation timescales remain $\Delta F$ independent in agreement with linear response ($\tau_F=0.029(2)s,\tau_I=1(1)s,\tau_S=41(3)s$) but decrease relative to the untreated T case ($\tau_F=0.30(5)s,\tau_I=4(1),\tau_S=70(8)s$) (Figure \ref{Fig:S8}c). Table \ref{table:tableviscoS1} of the Supplemental material shows the recovery force, amplitudes, and relaxation times for the treated and untreated cases. This finding is consistent with the observation of faster extrusion of RBC tethers in the absence of ATP. \cite{borghi2007tether} and the substantially lower entropy production of passivated RBCs \cite{di2023variance}. From Eqs. \ref{eq:DFr},\ref{eq:Ai},\ref{eq:Taui} we have extracted the viscoleastic parameters (Table \ref{table:tableviscoNG}). We find that the stiffnesses increase and the friction coefficients decrease relative to the untreated case, leading to RBC rigidification consistently with previous findings \cite{betz2009atp}. 

\begin{figure}[h]
\centering
\includegraphics[width=1.0\linewidth]{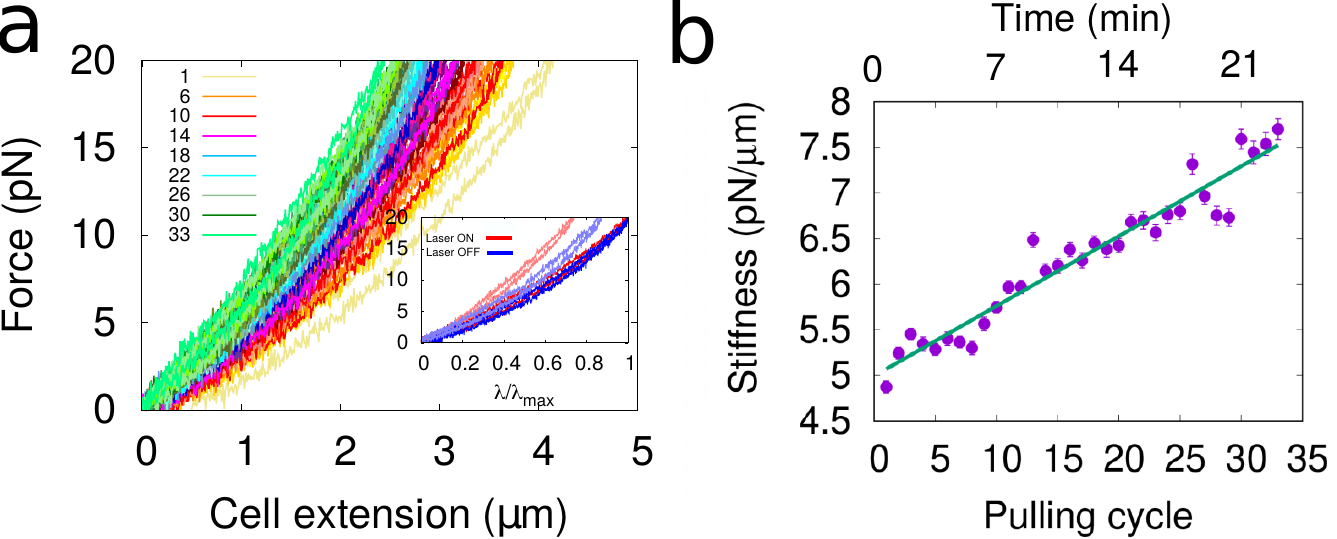}
\caption[Laser effect in RBC.]{\textbf{Laser effect in RBC.} (\textbf{a}) Force-extension curves (FECs) of an RBC over time. Each color represents a different pulling cycle and the index on the legend is the cycle number. Inset: FEC of an RBC with and without the laser effect. Dark colors represent the second pulling cycle, and light colors represent another pulling cycle after 45 minutes. Stiffening is apparent in the laser ON configuration. (\textbf{b}) RBC's stiffness was obtained by fitting the FEC of (a) between 7.5pN and 12.5pN. Results are shown for a representative RBC and are reproducible over different RBCs.}
\label{Fig:7}
\end{figure}

\begin{table}[htp]
\centering
\begin{tabular}{c|ccccc}
${\rm pN}$/$\mu{\rm m}$   & $k_{RBC}$  &  $k_{||}$ & $k_F$ & $k_I$ & $k_S$ \\ 
\hline

\textbf{T} & 6(2)  & 8.9(4)  &  3.7(6) & 2.6(3) & 2.6(3)  \\
\hline\hline
${\rm pN}\cdot {\rm s/\mu m}$ &  $\gamma_F$ & $\gamma_I$   & $\gamma_s$   \\ 
\hline
\textbf{T} & 0.10(2)  & 2.5(3) & 101(10) & \\
 \hline
\end{tabular}
\bigskip
  \caption{Viscoelastic model parameters for treated RBCs with tether (T). Parameters are obtained by fitting the viscoelastic model Eqs. \ref{eq:DFr},\ref{eq:Ai},\ref{eq:Taui} to the measured recovery force, amplitudes, and relaxation times. Table with stiffnesses and friction coefficients for treated and untreated cases can be found in Table \ref{table:tableviscoS2} of the supplemental material.} \label{table:tableviscoNG}
\end{table} 

In another set of experiments, we studied the irradiation effect of the laser on the RBC at normal power ($\lambda$=845nm, $P\sim$80mW). This kind of damage is relevant in radiation therapies for cancer treatment \cite{inanc2021quantifying}. In movie S3, Sec.\ref{Movie}, we show the effect of the infrared light on an RBC directly captured in the optical trap. Upon direct illumination, the RBC collapses after roughly 30 seconds. We have investigated the effect of the optical trap in the less invasive configuration of the pulling experiments (Figure\ref{Fig:1}a,d) where the RBC approaches the focus of the optical trap at low forces. We have carried out the experiments shown in Figure \ref{Fig:1}b for untethered RBCs to explore the cumulative effect of light on the RBC after repeated pulling cycles. As shown in Figure \ref{Fig:7}a, RBC elongation shortens at every cycle showing that RBCs rigidify upon light exposure at a rate of $0.077(4)$pN/$\mu$m per pulling cycle approximately doubling every $\sim 30$min, Figure \ref{Fig:7}b. Irradiation effects mainly occur at low forces during a pulling cycle, being residual when the RBC is pulled far from the light trap. To confirm this we have implemented a pulling protocol where we wait at zero force for a given time in two conditions: laser switched on at the normal power or at a very low trapping power ($P\sim$16mW). For the lower power the shortening is less apparent demonstrating the direct effect of light on RBC stiffening (Figure \ref{Fig:7}a, inset). A similar effect has been observed in RBC stretching experiments using electric fields \cite{qiang2019mechanical}. It has also been reported that repeated stretching cycles might induce ATP loss from the RBC, which has membrane channels that let ATP go out when the cell is stretched \cite{gov2005red}. Another effect that could contribute is vesiculation: as the RBC loses ATP and is mechanically deformed, the membrane area can decrease \cite{gov2009cytoskeletal}. In fact, upon stretching without the laser effect (Figure \ref{Fig:7}a (inset) blue colors-Laser OFF), a shortening of the extension is observed but of lesser magnitude than that observed with normal laser ON conditions. We may conclude that laser illumination contributes to stiffening the RBC.

\section*{Discussion and Conclusions}
%
RBCs continuously experience mechanical stress in vivo, showing the importance of investigating their mechanical response using biophysical techniques. We have investigated the mechanical deformability and viscoelastic response of RBCs in pulling and force-relaxation experiments in different types of extension-jump protocols. We have found that RBCs exhibit a linear viscoelastic behavior with a triple-exponential relaxation function and three well-separated timescales $\tau_F\sim 0.01-0.1$s (fast), $\tau_I\sim 4$s (intermediate),$\tau_S\sim 70$s (slow) spanning four decades in time, 0.01-100s. Most of the literature on RBCs is focused on the untethered (U) case where there is no tether formation (Figure \ref{Fig:1}a), and a single exponential is observed with a relaxation time of 0.1-0.6 s \cite{cranston1984plasmodium, hochmuth1979red, henon1999new, artmann1995microscopic, qiang2019mechanical, bronkhorst1995new, dao2003mechanics, chien1978theoretical}. Importantly, this timescale is compatible with the value of $\tau_F$ observed in the  case where a tether (T) is formed (Figure \ref{Fig:1}d). The fast timescale $\tau_F$ also falls in the range of theoretical predictions (0.01-0.1s) \cite{gov2007active,ben2011effective} and ultrafast microscopy experiments \cite{rodriguez2015direct}. Relaxation times around the intermediate timescale (1.27-1.5 s) have been also observed, which can be attributed to either the application of a large force (500pN) or mechanical fatigue \cite{sorkin2018probing, qiang2019mechanical}. Moreover, a timescale 10-100 times larger than the fast timescale $\tau_F$ has been attributed to spectrin network reorganization \cite{ben2011effective}.  Relaxation studies on the T case have primarily focused on the length and extrusion velocity of the tether rather than the relaxation times \cite{borghi2007tether}. Consequently, it becomes challenging to establish a direct comparison between timescales. However, recent investigations have examined the extrusion ($\Delta F>0$) of membrane tubes, and they have reported relaxation times of 0.3-0.7 s \cite{paraschiv2021influence}. These findings, which align with the relaxation times reported for the U case, were obtained based on adherent cell experimental data. Surprisingly, no literature reports a significantly longer timescale on tens of seconds, which would be compatible with the slow timescale $\tau_s\sim 70$s, for either the U or T cases. Here, we hypothesize that the relaxation times reported in the literature combine the fast and intermediate timescales, while $\tau_F$ differs in the T and U cases. As a result, both $\tau_F$ and $\tau_I$ can be linked to the relaxation of the cell membrane, which, according to previous studies, takes place within a few seconds \cite{cranston1984plasmodium, hochmuth1979red, henon1999new, artmann1995microscopic, qiang2019mechanical, bronkhorst1995new, dao2003mechanics, chien1978theoretical,sorkin2018probing}. To the best of our knowledge, a $\tau_S$ on the order of tens of seconds has never been reported, probably because relaxation experiments have never reached the timescale of minutes and the slow decay amplitude ($A_S$) of the relaxation function is masked by the larger amplitudes $A_F,A_I$ of the fast and intermediate processes (Figure \ref{Fig:2}c). What is the origin of the largest timescale $\tau_S$? Dynamical scaling implies that large timescales correspond to larger lengthscales, suggesting that a $\tau_S$ of 70s must arise from a collective process spanning the RBC's body. It is worth noting that other studies of RBC relaxation with optical tweezers have already fitted force-relaxation curves to multiple timescales \cite{yoon2008nonlinear} but did not find three distinct timescales. Furthermore, in systems such as giant vesicles, multiple relaxation timescales have been reported using electro-deformation, in the range of tens of milliseconds to seconds, but not tens of seconds \cite{zhou2011stretching, riske2005electro}. 

A viscoelastic model shows that the stiffness of the RBC measured in pulling experiments, $k_{RBC}$, is comparable to the total stiffness of the three viscoelastic processes in a parallel configuration ($k_{||}=k_F+k_I+k_S$, Fig.\ref{Fig:5}). Interestingly, the friction coefficients $\gamma_I,\gamma_S$ are at least a hundred times larger than $\gamma_F$, with and without a tether (Table \ref{table:tablevisco}). We interpret the fast process ($\tau_F$) as due to the local dynamics of the membrane, the intermediate process ($\tau_I$) as due to the membrane-cortex interaction, and the slow process ($\tau_S$) to the high cytosol viscosity (>1 Pa$\cdot$s). Our interpretation agrees with the observation that the timescale $\tau_F$ depends on the local geometry of the deformed lipid bilayer between RBC and bead, which is different for the T and U cases. On the other hand, $\tau_I$ and $\tau_S$ do not change between the T and U cases because the membrane-cortex and the cytosol deform similarly in the force-relaxation experiments. 

Upon depleting glucose (Figure \ref{Fig:6}), the three timescales decrease, with the stiffness parameters $k_F,k_I,k_S$ increasing and the friction coefficients $\gamma_F,\gamma_I,\gamma_S$ decreasing. Therefore glucose depletion solidifies the RBC by increasing the elastic relative to the viscous response. This result is compatible with the general observation reported in the literature that environmental damage and stress rigidify RBCs. In particular, we have also  demonstrated that under light illumination, RBCs become stiffer at a rate that increases with the laser power (Figure \ref{Fig:7}).  

Overall, our study reveals structured viscoelastic dynamics. The observed triple-exponential behavior with three separated timescales lies in between the pure exponential relaxation observed in linear and two-level systems \cite{del1995monotonic} and the more complex stretched-exponential relaxation observed in polymers, glassy and amorphous matter \cite{yu2015stretched}. We may call this discrete-stretched exponential behavior where the finite number of relaxational timescales is consequential to the organized cell structure. One might ask whether there are other timescales beyond the measured time range in this study ($\sim$ 5 minutes). Answering this question would require longer experiments in the timescale of hours. Further evidence of structured viscoelastic dynamics is the dynamical scaling observed between the three timescales (Fig.\ref{Fig:4}d). Although they vary over several orders of magnitude across the RBC population, they mutually scale following a power law, a consequence of the underlying RBC architecture. Future studies should investigate the effect of modifying the lipid bilayer composition and drugs disruptive of the cortex to clarify the current interpretation of the three timescales. Moreover, it would be interesting to explore these relaxational phenomena in other types of RBC (e.g. sickle anemia), lymphocytes, and the effect of changing temperature on RBC dynamics.

\section*{Author Contributions}
M.G. collected and cured the data, wrote the software for data analysis and took care of visualization, and analyzed the data. F.R. administered the project and supervised the research. M.G and F.R. wrote the original draft. G.B, R.S and G.W discussed the results and implications of the methodology and commented on the manuscript.

\section*{Acknowledgments}
M.G. and F.R. are supported by the Spanish Research Council Grant PID2019-111148GB-100. F.R. is supported by the ICREA Academia Prize 2018.  R.S. acknowledges support through Human Frontier Science Program (HFSP) postdoctoral fellowship LT000419/2015. This project has received funding from the European Research Council (ERC) under the European Union’s Horizon 2020 research and innovation program (grant agreement No. [883240]) to G.W. Moreover, G.W. likes to acknowledge support by the Netherlands Organisation for Scientific Research (NWO/OCW), as part of the BaSyC Gravitation program. We are grateful to N. Gov and A. Hernandez-Machado for a critical reading of the manuscript.


\newpage
\renewcommand{\thefigure}{S\arabic{figure}}
\renewcommand\theequation{S\arabic{equation}}
\renewcommand{\thesection}{S\arabic{section}}
\setcounter{figure}{0}
\setcounter{equation}{0}
\setcounter{page}{1}
\let\oldthetable\thetable
\renewcommand{\thetable}{S\oldthetable}
\setcounter{table}{0}

\begin{center}
{\LARGE Supplemental Material for}\\
\vspace{0.4cm}
\textbf{\Large Viscoelastic phenotyping of red blood cells}
\end{center}
\vspace{1cm}

\noindent
{\bf This PDF contains}:\\\\
Supplemental Text\\
Figs. S1 to S8\\
Tables S1 and S2 \\
Movies S1 and S3\\

{\hypersetup{linkcolor=black}
\tableofcontents
}

\clearpage

\section{Pulling experiments}\label{sec:S1Pulling} 

 \begin{figure}[H]
 \centering
 \includegraphics[width=0.5\linewidth]{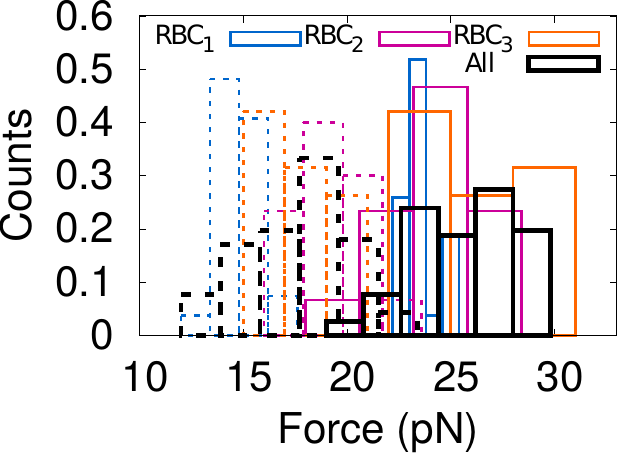}
 \caption[Force histograms of extrusion/absorption of tether in pulling experiments]{\textbf{Force histograms of extrusion/absorption of tether in pulling experiments.} In continuous lines, extrusion force histograms of three different RBCs (colors) together with the total average (black). In dashed lines, absorption force histograms of three different RBCs (colors) together with the total average (black).}
 \label{Fig:S1}
 \end{figure}

 \begin{figure}[H]
 \centering
 \includegraphics[width=1.0\linewidth]{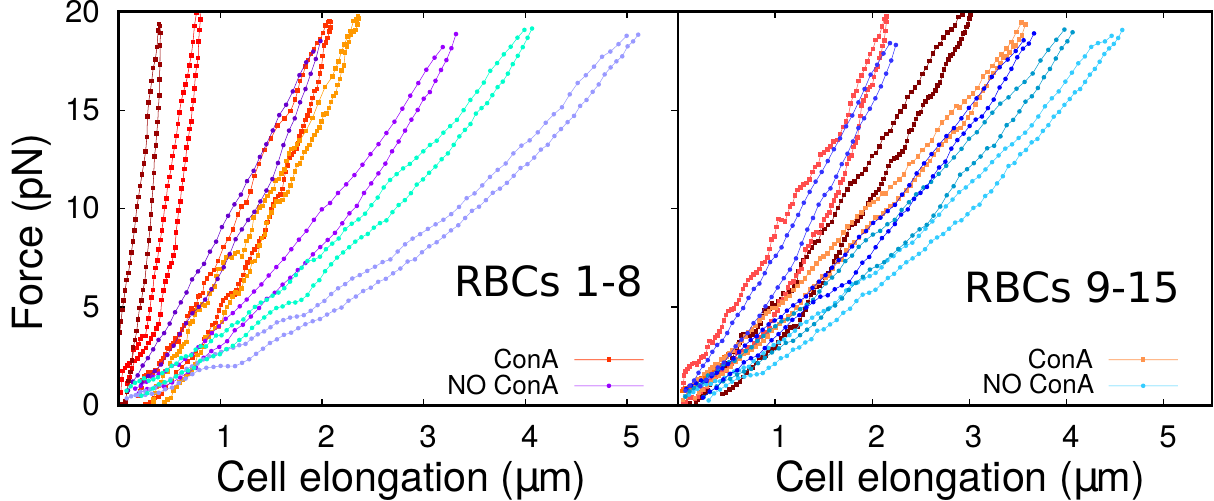}
 \caption[Concanavilin A coating in pulling experiments]{\textbf{Concanavilin A coating in pulling experiments.} Force respect to cell elongation for different RBCs. Blue colors represent force extension curves (FEC without ConA coating and red FEC represent RBC with ConA coating. The FEC have been divided in two panels indifferently in order to facilitate their visualization.}
 \label{Fig:S2}
 \end{figure}

 \section{Relaxation experiments}\label{sec:S2Relax}

 \begin{figure}[H]
\centering
\includegraphics[width=1\linewidth]{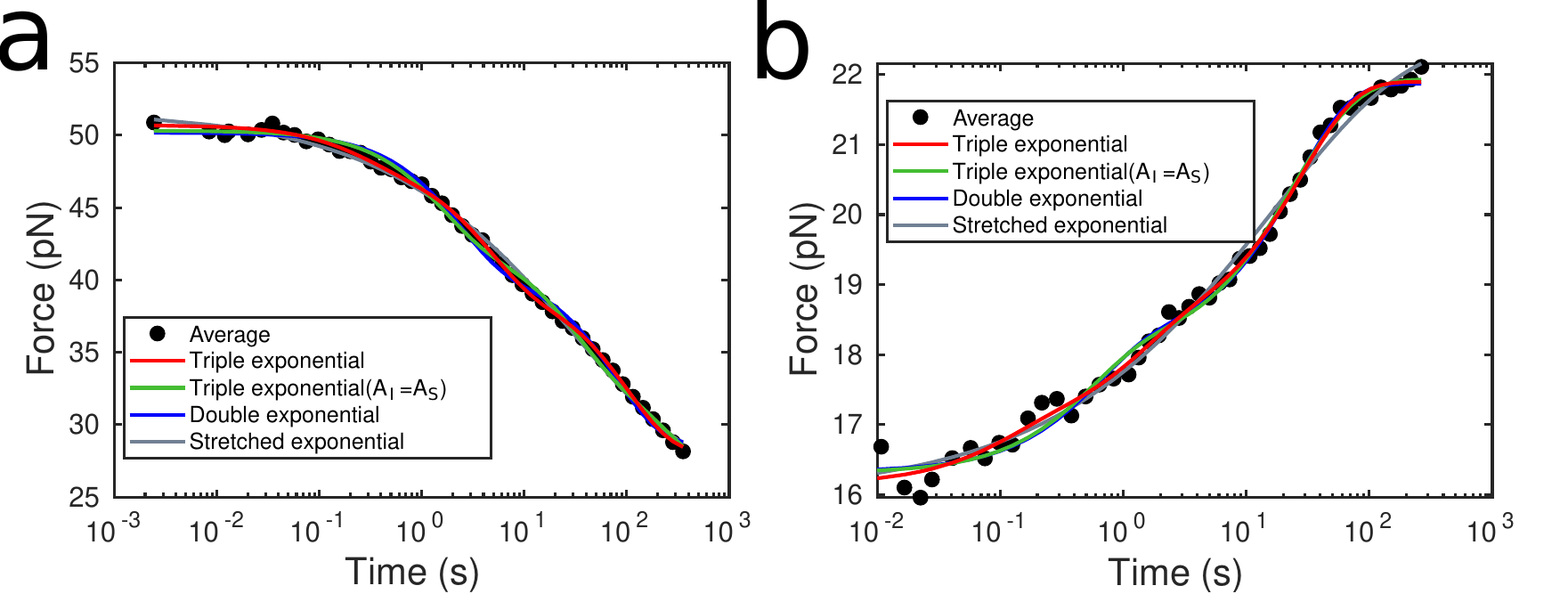}
\caption[Force relaxation curve (FRC) with different fits]{ \textbf{Force relaxation curve (FRC) with different fits.} Average of the FRC (black points) with fits to triple exponential (red), triple exponential imposing intermediate amplitude equal to slow one (green), double exponential (blue), and stretched exponential (grey). FRCs shown for  (\textbf{a})  Trotter(+) and (\textbf{b}) Ladder(-). }
\label{Fig:S3}
\end{figure}

 \begin{figure}[H]
\centering
\includegraphics[width=0.55\linewidth]{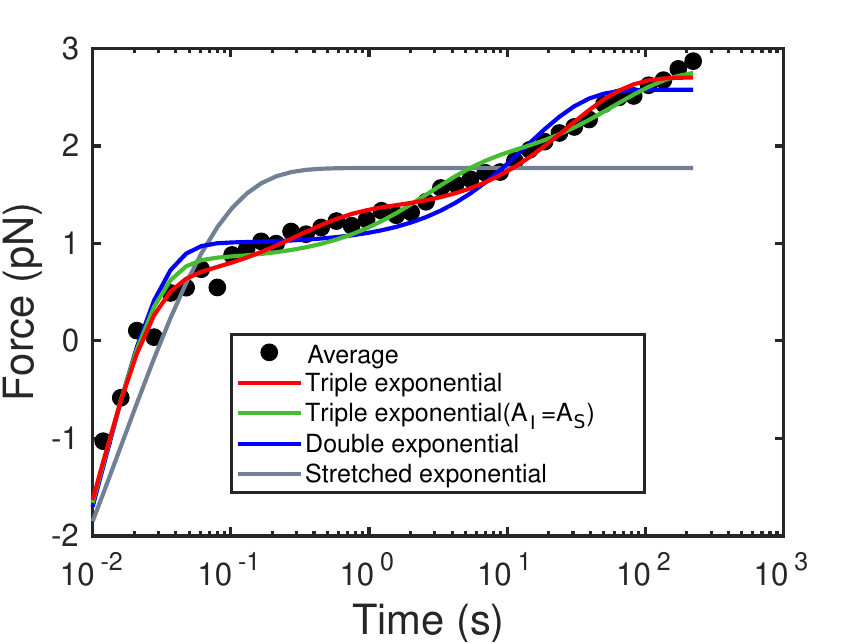}
\caption[Relaxation of a partially ATP-depleted RBC]{\textbf{Force relaxation curve of an ATP-depleted RBC.} Average of the FRC (black points) with fits to triple exponential (red), triple exponential imposing intermediate amplitude equal to slow one (green), double exponential (blue), and stretched exponential (grey). FRC shown for Trotter(-) }
\label{Fig:S7}
\end{figure}
 
 \begin{figure}[H]
\centering
\includegraphics[width=0.55\linewidth]{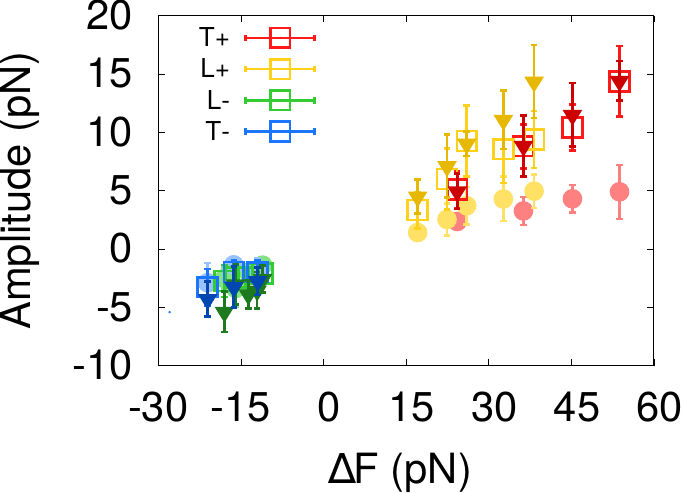}
\caption[Amplitudes for the different protocols]{\textbf{Amplitudes for the different protocols.} Amplitudes as a function of $\Delta F$ for Trotter(+) in red, Ladder(+) in yellow, Trotter(-) in blue, and Ladder(-) in green. Circles represent the fast amplitude, squares represent the intermediate amplitude, and triangles represent the slow amplitude.}
\label{Fig:S4}
\end{figure}

\subsection{Viscoelastic model derivation}\label{sec:S21Model} 

\begin{figure}[H]
 \centering
 \includegraphics[width=0.5\linewidth]{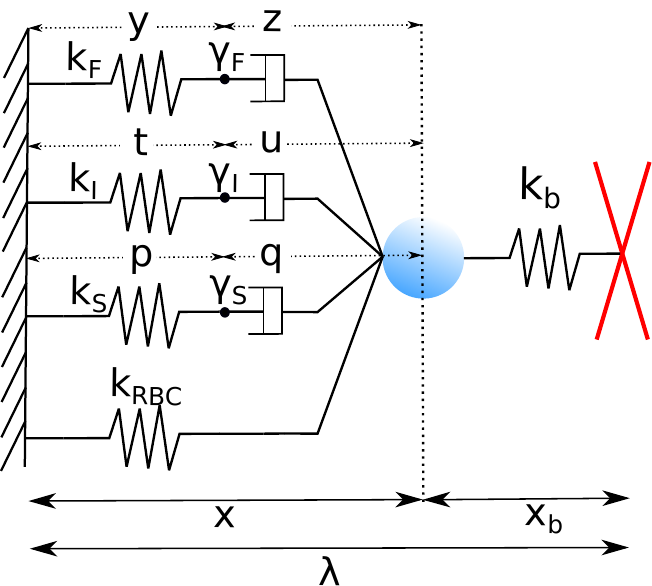}
 \caption[Schematics of the RBC]{\textbf{RBC viscoelastic model based on parallel combination of Maxwell units.}}
 \label{Fig:S5}
 \end{figure}

We model the RBC by connecting in parallel three Maxwell units (a linear spring and a dashpot in series) with a single linear spring. This way, the viscoelasticity observed in the relaxation experiments, characterized by three exponential processes and a recovery force, is reproduced. In Figure \ref{Fig:S5}, an schematics of the RBC model in the LOT setup is shown. 

In this configuration $\lambda$ is defined as the total distance that results to combined the RBC extension $x$ and the trap extension $x_b$,

  \begin{equation}
      \lambda = x + x_b
  \end{equation}

At the same time, $x$ is the sum between the extension of the spring and the dashpot connected in series for each of the three processes,

    \begin{equation}
      x = y+z = t + u = p +q
      \label{eq:dist}
  \end{equation}

The force exerted to each Maxwell unit is expressed as,

    \begin{equation}
      f_F = k_F y = \gamma _F \Dot{z}
      \label{eq:f1}
  \end{equation}

      \begin{equation}
      f_I = k_I t = \gamma _I \Dot{u}
  \end{equation}

      \begin{equation}
      f_S = k_S p = \gamma _S \Dot{q}
  \end{equation}

where $k_F$, $k_I$ and $k_S$ are the springs' stiffness and $\gamma _F$, $\gamma _I$ and $\gamma _S$ are the dashpots' friction coefficients of the fast, intermediate ans slow processes, and $\Dot{z}$, $\Dot{u}$ and $\Dot{q}$ are the derivatives respect to time of the dashpots extensions. 
With respect to the optical trap, the exerted force is the product between the trap stiffness $k_b$ and the trap extension $x_b$,
      \begin{equation}
      F = k_b x_b = k_b (\lambda -x)
      \label{eq:fOT}
  \end{equation}

Taking into account that the total force acting on the bead is zero, the force exerted by the optical trap has to be equal to the one exerted by the RBC,

\begin{equation}
    F=f_F+f_I+f_S+f_{RBC}
\label{eq:eqF}
\end{equation}

where, $f_{RBC}=k_{RBC}x$, and, therefore combining eqs. \ref{eq:fOT} and \ref{eq:eqF} , $x$ is expressed as,

        \begin{equation}
      x = -\frac{1}{(k_{RBC}+k_b)} (f_F + f_I + f_S - k_b\lambda)
      \label{eq:distF}
  \end{equation}

  From eq. \ref{eq:f1} we get the relation,

      \begin{equation}
      \Dot{y} + \Dot{z} = \frac{\Dot{f}_F}{k_F} + \frac{f_F}{\gamma _F} 
      \label{eq:yzpunt}
  \end{equation}

  Deriving eqs. \ref{eq:dist} and \ref{eq:distF} with respect to time and combining them with eq. \ref{eq:yzpunt}, assuming steady-state conditions, $\Dot{\lambda} =0$, we get for the fast process,
  
      \begin{equation}
      -\frac{1}{(k_{RBC}+k_b)} (\Dot{f}_F + \Dot{f}_I + \Dot{f}_S) = \frac{\Dot{f}_F}{k_F} + \frac{f_F}{\gamma _F} 
      \label{eq:Tau1}
  \end{equation}

 Like \ref{eq:Tau1}, we obtain two other equivalent equations for the intermediate and slow processes by replacing the rhs of eq. \ref{eq:Tau1} with the slow and intermediate processes ($f_F \to f_S, f_I$). By equating the rhs of these three expressions, we rewrite $\Dot{f}_I$ in terms of $f_I$,$\Dot{f}_F$ and $f_F$ and $\Dot{f}_S$ in terms of $f_S$,$\Dot{f}_F$and$f_F$, and obtain two independent equations,

        \begin{equation}
      \Dot{f}_I = \frac{k_I}{k_F}\Dot{f}_F + \frac{k_I}{\gamma_F}f_F - \frac{k_I}{\gamma_I}f_I  
      \label{eq:f1dot}
  \end{equation}

          \begin{equation}
      \Dot{f}_S = \frac{k_S}{k_F}\Dot{f}_F + \frac{k_S}{\gamma_F}f_F - \frac{k_S}{\gamma_S}f_S  
      \label{eq:f3dot}
  \end{equation}

Substituting eqs. \ref{eq:f1dot} and \ref{eq:f3dot} in the lhs of eq. \ref{eq:Tau1}, the latter is rewritten as,



        \begin{equation}
      -\frac{\Dot{f}_F}{k_F} (k_F + k_I + k_S + k_{RBC}+k_b) = \frac{f_F}{\gamma_F} ( k_F + k_S + k_{RBC}+k_b) - f_I \frac{k_I}{\gamma_I} - f_S \frac{k_S}{\gamma_S} 
      \label{eq:Tau4}
  \end{equation}

  To simplify we define the total stiffness as, $k_T = k_F + k_I + k_S + k_{RBC}+k_b$. From eq. \ref{eq:Tau4} we obtain two equivalent but independent equations by permuting the indices $F \leftrightarrow I \leftrightarrow S$. Therefore, we get a generalized equation as,

  \begin{equation}
      -\frac{\Dot{f}_i}{k_i} (k_T) = \frac{f_i}{\gamma_i} ( k_T-k_i) - f_j \frac{k_j}{\gamma_j} - f_k \frac{k_k}{\gamma_k}   
      \label{eq:Taufinal}
  \end{equation}

where $i,j,k$ are different indices for the fast, intermediate and slow processes. To solve eq. \ref{eq:Taufinal} we assume well-separated time scales. For the fast process, we assume that in the slow and intermediate processes the forces have not yet relaxed and therefore $\Dot{f}_I=\Dot{f}_S=0$ and hence $f_I,f_S$ can be taken as constants,

  \begin{equation}
      -\frac{\Dot{f}_F}{k_F} (k_T) = \frac{f_F}{\gamma_F} ( k_T-k_F) - f_I \frac{k_I}{\gamma_I} - f_S \frac{k_S}{\gamma_S}   
      \label{eq:TaufinalF}
  \end{equation}

A similar argument for the intermediate and the slow processes yield to a generalize expression for the relaxation time,
  
  \begin{equation}
      \tau_i = \frac{\gamma_i k_T}{k_i( k_T-k_i)} 
      \label{eq:TaufinalF2}
  \end{equation}

The force amplitude of each spring-dashpot process is defined as,

    \begin{equation}
        f_i = k_i \Delta x
    \label{eq:Ampi}
  \end{equation}
  
Combining eqs. \ref{eq:Ampi} and \ref{eq:eqF}, we can express the displacement in terms of the force jump, $\Delta F$ as,
 
  \begin{equation}
       \Delta x = \frac{\Delta F}{k_{RBC}+k_{||}} 
      \label{eq:Feq2}
  \end{equation}

where a $k_{||}= k_F + k_I + k_S$. Therefore, substituting eq. \ref{eq:Feq2} in eq. \ref{eq:Ampi}, we get the generalize expression for the amplitudes in terms of $\Delta F$,

    \begin{equation}
       f_i = \frac{k_i}{k_{RBC}+k_{||}} \Delta F
      \label{eq:Dx}
  \end{equation}
  
We define the recovery force $\Delta F_R$ as,
    \begin{equation}
       \Delta F_R = \Delta F - \Delta F'
      \label{eq:DFR}
  \end{equation}

where $\Delta F$ is the force jump and $\Delta F'$ is the difference between the force after the five minutes relaxation, $F_f$, and the force before the jump, $F_i$ ($\Delta F' = F_f - F_i $) which can also be expressed as, 

    \begin{equation}
       \Delta F' = k_{RBC} \Delta x' 
      \label{eq:DFp2}
  \end{equation}

Then, as $\Delta F' = k_b (\Delta \lambda - \Delta x')$, we get the corresponding displacement $\Delta x'$ in terms of $\Delta \lambda$,

    \begin{equation}
       \Delta x' = \frac{k_b}{k_{RBC}+k_b} \Delta \lambda 
      \label{eq:Dxp}
  \end{equation}

  Substituting eq. \ref{eq:Dxp} in eq. \ref{eq:DFp2}, we get,

      \begin{equation}
       \Delta F' = \frac{k_{RBC}k_b}{k_{RBC}+k_b} \Delta \lambda 
      \label{eq:DFp3}
  \end{equation}

  To obtain $\Delta \lambda (\Delta F)$, we substitute eq. \ref{eq:Feq2} in $\Delta F = k_b (\Delta \lambda - \Delta x)$, obtained from the rhs of eq. \ref{eq:fOT},

    \begin{equation}
       \Delta \lambda = \frac{k_{RBC}+k_{||}+k_b}{k_b(k_{RBC}+k_{||})} \Delta F
      \label{eq:DFp4}
  \end{equation}

  Substituting eq. \ref{eq:DFp4} in eq. \ref{eq:DFp3}, we obtain $\Delta F'$ in terms of $\Delta F$,

      \begin{equation}
       \Delta F' = \frac{k_T}{(k_b(k_T-k_b))} \frac{k_{RBC}k_b}{(k_{RBC}+k_b)} \Delta F
      \label{eq:DFp5}
  \end{equation}

  Then the recovery force can be expressed as,

     \begin{equation}
       \Delta F_R = \Delta F (1 - \frac{k_T}{(k_b(k_T-k_b))} \frac{k_{RBC}k_b}{(k_{RBC}+k_b)})
      \label{eq:DFR2}
  \end{equation} 

Simplifying eq. \ref{eq:DFR2}, we obtain the final expression for the recovery force,

       \begin{equation}
       \Delta F_R =  \frac{\Delta F}{(1+\frac{k_{RBC}}{k_b})(1+\frac{k_{RBC}}{k_{||}})}
      \label{eq:DFRf}
  \end{equation}

 \begin{figure}[H]
\centering
\includegraphics[width=1\linewidth]{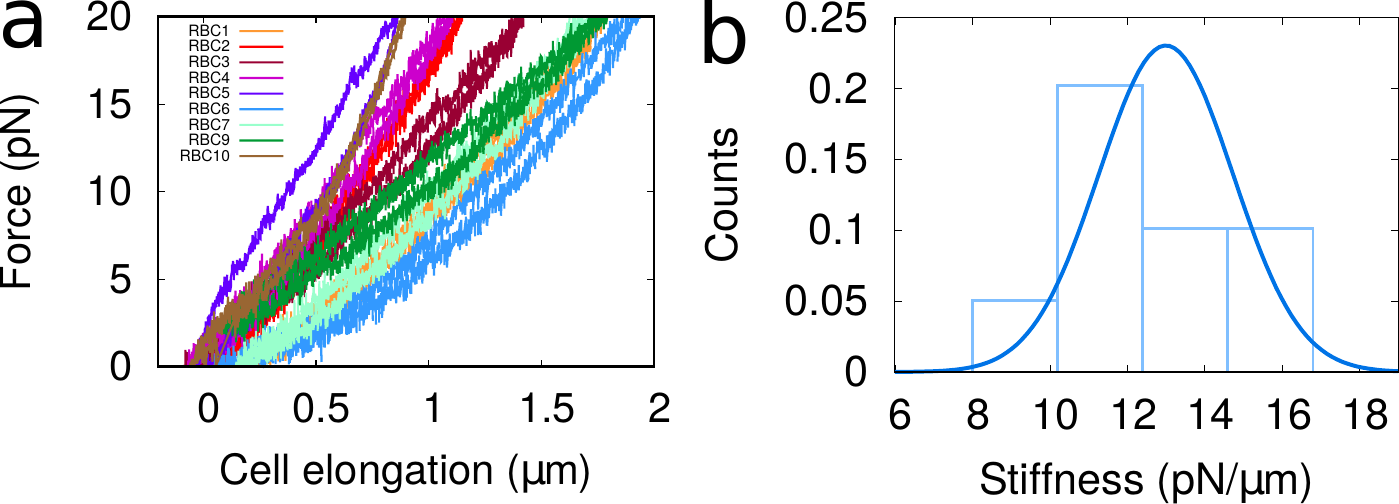}
\caption[ATP-depleted RBC pulling experiments]{\textbf{ATP-depleted RBC pulling experiments.} (a) Force extension curves of 10 different ATP-depleted RBCs. (b) Histogram of RBCs stiffness. The cell stiffness is obtained by performing a linear fit between 7.5pN and 12.5pN in the FEC. }
\label{Fig:S6}
\end{figure}

\begin{table}[htp]
\centering
\begin{tabular}{c|cccccc}

   & $A_F/\Delta F$ & $A_I/\Delta F$ & $A_S/\Delta F$  &  $\tau_F$ (s) & $\tau_I$ (s)   & $\tau_S$ (s)   \\ 
\hline
\textbf{Healthy Untethered}  &  0.11(1) & 0.15(2) & 0.15(2) & 0.020(5)  & 4(1) & 70(8)  \\
\textbf{Healthy Tethered} &   0.11(1) & 0.27(3) & 0.27(3)  & 0.020(5)/ 0.30(5) & 4(1) & 70(8)  \\
\textbf{Treated Tethered} &   0.31(3) & 0.20(3) & 0.20(3)  &   0.029(2) & 1(1) & 41(3)  \\
\hline
\end{tabular}
\bigskip
  \caption{Fitting parameters of the triple exponential viscoelastic model. Parameters are obtained by fitting Eqs. \ref{eq:DFr},\ref{eq:Ai},\ref{eq:Taui} to the measured recovery force, amplitudes and relaxation times.}
     \label{table:tableviscoS1}
\end{table}

\begin{table}[htp]
\centering
\begin{tabular}{c|ccccc|ccc}
  & \multicolumn{5}{c|}{${\rm pN}/ \mu{\rm m}$} & \multicolumn{3}{c} {${\rm pN}\cdot {\rm s/\mu m}$}  \\
  \hline
   & $k_{RBC}$  &  $k_{||}$ & $k_F$ & $k_I$ & $k_S$ &  $\gamma_F$  & $\gamma_I$  & $\gamma_s$   \\  
\hline
$\textbf{Healthy Untethered}$ & 6.1(13)  & 5.3(1)  &  1.3(1) & 2.0(1) & 2.0(1) & 0.025(3)  & 7.8(6) & 136(12) \\
$\textbf{Healthy Tethered}$ & 2.7(6)  & 6.4(4)  &  1.0(1) & 2.7(5) & 2.7(5) & 0.29(2)  & 10.2(7) & 179(13) \\
$\textbf{Treated Untethered}$ & 13(3)  & -  &  - & - & - & -  & - & - \\
$\textbf{Treated Tethered}$ & 6(2)  & 8.9(4)  &  3.7(6) & 2.6(3) & 2.6(3)  & 0.10(2)  & 2.5(3) & 101(10) \\
\hline
\end{tabular}
\bigskip
  \caption{Stiffnesses and friction coefficients obtained by fitting Eqs. \ref{eq:TaufinalF2},\ref{eq:Dx},\ref{eq:DFRf}.}
     \label{table:tableviscoS2}
\end{table}

 \begin{figure}[H]
\centering
\includegraphics[width=1\linewidth]{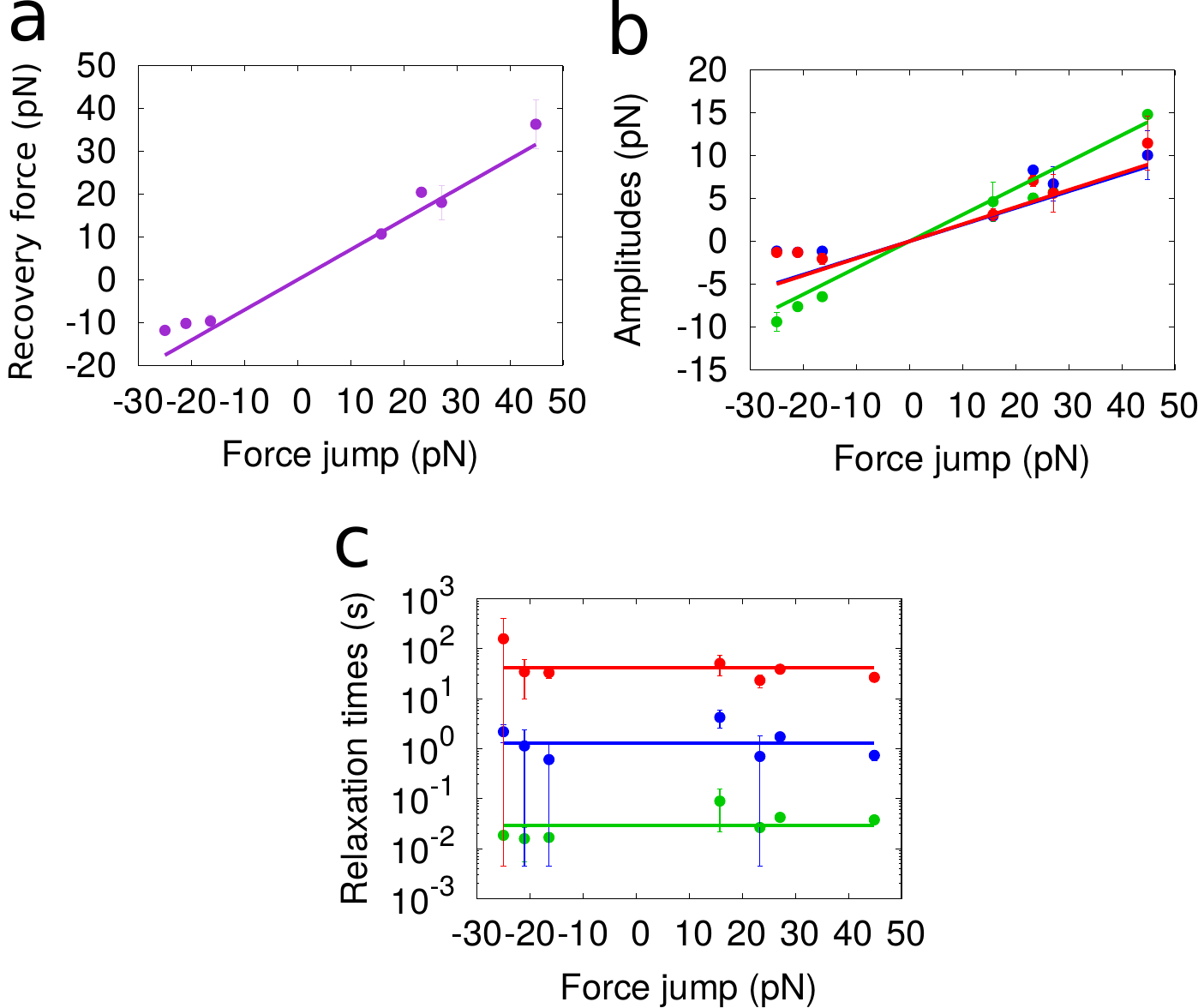}
\caption[ATP-depleted RBC relaxation parameters]{\textbf{ATP-depleted RBC relaxation parameters.} (a) Recovery force versus jump force with the corresponding linear fit. (b) Amplitudes versus jump force with their corresponding linear fits. (c) Relaxation times versus jump force with their corresponding fits to a constant value. }
\label{Fig:S8}
\end{figure}

\subsection*{Movies}\label{Movie}
Movie S1: \textbf{Formation of a tether in a tensional relaxational experiment.} A very thin tether hardly observable is former upon increasing the force ($\Delta F>0$) and stretching the RBC after sudden movement of the pipette downwards at time 3 seconds after starting the movie.

\noindent Movie S2: \textbf{Formation of a tether in a compressional relaxational experiment.} A thicker observable tether is former upon reducing the force ($\Delta F<0$) and compressing the RBC after sudden movement of the pipette upwards at time 3 seconds after starting the movie.

\noindent Movie S3: \textbf{Effect of laser illumination on an optically trapped RBC.} It can be observed that the RBC membrane breaks and releases body content. RBC material is observed to be released at time 29 seconds after starting the movie.



\end{document}